\documentclass[%
 reprint,amsmath,amssymb,
 aps
]{revtex4-2}
\usepackage{graphicx}
\usepackage{dcolumn}
\usepackage{bm} 
\usepackage{amsmath}
\usepackage{hyperref}
 \usepackage{bigints}   
\usepackage{graphicx}
\usepackage{caption}
\usepackage{subcaption}
\usepackage{setspace}
\usepackage{caption}
\usepackage{color}
\def \O{\mathcal{O}}
\def \x{\vec{x}}
\def \k{\vec{k}}
\def \bx{\mathbf{x}}
\def \bk{\mathbf{k}}
\def \+{\Delta}
\def \-{{d-\Delta}}
\def \d{\dagger}
\def \beq{\begin{equation}}
\def \eeq{\end{equation}}
\def \Z{{\mathbb Z}}
 
\begin{document}

\title{Bulk Reconstruction in De Sitter Spacetime}

\author{Arundhati Goldar }
\email{d21086@students.iitmandi.ac.in}
\author{Nirmalya Kajuri}%
 \email{nirmalya@iitmandi.ac.in}
\affiliation{%
School of Physical Sciences, IIT Mandi,\\ Himachal Pradesh, India}%

\date{\today}

\begin{abstract}
The bulk reconstruction program involves expressing local bulk fields as non-local operators on the boundary. It was initiated in the context of AdS/CFT correspondence. Attempts to extend it to de Sitter have been successful for heavy(principal series) scalar fields. For other fields, the construction ran into issues. In particular, divergences were found to appear for higher spin fields. In this paper, we resolve these issues and obtain boundary representations for scalars of all masses as well as higher spin fields. We trace the origin of the previously discovered divergences and show that the smearing function becomes distributional for certain values of mass, spin and dimension. We also extend the construction from Bunch-Davies vacuum to all $\alpha$-vacua. 
\end{abstract}
 
\maketitle

\section{Introduction}

The bulk reconstruction program has its origin in the context of AdS/CFT correspondence. While the AdS/CFT correspondence\cite{Maldacena:1997re,Gubser:1998bc,Witten:1998qj} tells us of the equivalence between quantum gravity in an asymptotically Anti-de Sitter spacetime and a conformal field theory living on its boundary, it does not provide an explicit map between every bulk and boundary observable.

The bulk reconstruction program aims to find `boundary representations' of bulk fields. A boundary representation of a local bulk field is a non-local CFT operator whose correlation functions in the boundary theory replicate the correlation functions of the original bulk field. Boundary representations have been constructed for different fields and backgrounds\cite{Dobrev:1998md, Bena:1999jv,Hamilton:2005ju,Hamilton:2006az,Hamilton:2006fh,Heemskerk:2012np,Papadodimas:2012aq,Kabat:2011rz,Heemskerk:2012mn, Kabat:2012hp,Leichenauer:2013kaa,Sarkar:2014dma,Sarkar:2014jia,Guica:2014dfa,Roy:2015pga,Kabat:2016rsx,Kabat:2017mun,Kabat:2018pbj,Foit:2019nsr,Kajuri:2020bvi,Dey:2021vke,Bhattacharjee:2022ehq,Bhattacharjee:2023roq}. For a recent review, we refer to \cite{Kajuri:2020vxf}. 

Let us briefly outline the idea of bulk reconstruction, taking a free scalar field in AdS as an example. A free scalar in AdS has two independent solutions, but only one of them is normalizable. This mode falls off near the boundary as:
\begin{equation}
  \lim_{z\to 0}   \phi(z,\bx) \sim z^{-\Delta} \O_{\Delta}(\bx)
\end{equation}
where $z$ is a radial coordinate and $\bx=x_0$ are boundary coordinates. From AdS/CFT correspondence, $\O(\bx)$ are primary operators in the boundary conformal field theory of conformal dimension $\Delta= \frac{d}{2}+ \sqrt{\frac{d^2}{4}+m^2}$. Solving the bulk equation of motion and using the above equation, one can construct a boundary operator of the form$\phi$
\begin{equation}\label{smar}
\phi(z,x) =\int K(z,x;x^\prime)\O(x^\prime)   
\end{equation}
This is a non-local operator in the boundary field theory whose correlators are equal to the correlators of the bulk scalar. This is the boundary representation of the bulk scalar $\phi$. The function $K(z,x;x^\prime)$ is called the smearing function.

There are several reasons why one would want to carry out a similar construction for de Sitter. A key motivation for this comes from the possibility of a dS/CFT correspondence. It has long been conjectured that there should be (possibly non-unitary) CFT that is dual to an asymptotically de Sitter spacetime\cite{Strominger:2001pn}. While no such result has been established as yet, dS/CFT continues to be an intriguing area of research. If such a correspondence is established, then boundary representations of bulk fields would become useful in understanding the bulk-boundary map. One would be able to probe dS physics directly from the boundary CFT using boundary representations.

A second motivation comes from the cosmological bootstrap program\cite{Arkani-Hamed:2018kmz,Baumann:2019oyu,Sleight:2019hfp,Pajer:2020wxk,Baumann:2020dch,Hogervorst:2021uvp,DiPietro:2021sjt,Baumann:2022jpr}. Recall that for de Sitter space, there are two spacelike boundaries--the `late-time' and the `early time' boundary. The former is of cosmological interest as the point at which inflation ended. Correlation functions of bulk fields at this time are expected to show up in CMB observations. In the cosmological bootstrap program, conformal symmetry at the late-time boundary of dS is used (along with principles like unitarity) to constrain the late-time correlators of bulk fields.  Boundary representations of fields in dS analogous to \eqref{smar} would make it straightforward to map their late-time correlators to correlators at any prior time.

 The key difference between bulk reconstruction in dS and AdS is that both the independent solutions for a free scalar in dS are normalizable and have to be considered. So near the late-time boundary (we do not consider the early time boundary in this paper):
\begin{equation}
  \lim_{\eta\to 0}   \phi(\eta,x) \sim \eta^{\Delta} \O_{\Delta}(\x) + \phi(\eta,x) \sim \eta^{\Delta} \O_{d-\Delta}(\x)
\end{equation}
Here $\eta$ is the time coordinate and $\x$ are the spacelike coordinates.
The $\O_y(\x)$ are the boundary fields transform under de Sitter isometries in the same way as a conformal primary operator of dimension $y$ transforms under conformal transformations. If  the same steps as in AdS can be completed, one would obtain a boundary representation of the form: 
\begin{align}\notag
\phi(\eta,x) \sim &\int  K_\+(\eta,x;x^\prime)\O_\+(x^\prime) \\ + &\int K_\-(\eta,x;x^\prime)\O_\-(x^\prime) 
\end{align}

 Bulk reconstruction in a $d+1$-dimensional de Sitter spacetime was carried out for scalar fields of mass $(m/H)^2>d^2/4$ in \cite{Xiao:2014uea}(see also \cite{Sarkar:2014dma,Sarkar:2014jia} for follow-up work and discussion on higher spin fields in dS, and \cite{Chatterjee:2015pha,Bhowmick:2018bmn,Doi:2024nty} for different approaches). The strategy used in \cite{Xiao:2014uea} was to read off the smearing functions from the Wightman function of a scalar in dS. As noted there, this strategy only applies to scalars fields belonging to the principal series. The validity of the results beyond this range was unclear.  
 
In particular, one ran into problems if one tried to apply the results thus obtained to scalars for which $\Delta$ is either zero or a negative integer. The smearing function diverges in this case. This case is of physical interest because higher spin fields in dS can be recast as a multiplet of exactly such scalars. 

Therefore, boundary representations for scalars of all masses and for higher spin fields remained an open problem. Further, boundary representations have only been constructed for the Bunch-Davies (or Euclidean) vacuum.
 
In our paper, we use a different approach to bulk reconstruction, the so-called mode sum approach\cite{Hamilton:2005ju}. Using this method, overcame the above-mentioned limitations and construct the smearing function for the most general case of a bosonic field of arbitrary spin in de Sitter. Further, we extend the construction of boundary representations to arbitrary $\alpha$-vacua.

We uncover several interesting features. First, we find that a different formula holds for the integer $\Delta$ case, which has the same functional form as the non-integer case, but not the divergences. Second, we find that all the smearing functions are given by Weber-Schafheitlin type integrals, which are well known in the mathematical literature. This integral only gives analytic expressions for a certain range of the parameters involved (in our cases, it translates to conditions on $\Delta$ and $d$). Beyond this range, smearing functions becomes distributional. This explains the appearance of divergences discussed above. However, useful (distributional) expressions are available even in this range.

In asymptotic AdS spacetimes, smearing functions are distributional in the presence of event horizons (black holes or AdS-Rindler charts)\cite{Morrison:2014jha}. Here we find that the smearing function can be distributional even in pure de Sitter. 

We also found that the smearing function vanishes in even dimensions, when $\Delta$ is an integer. It is unclear if this is of physical significance, or a limitation of the mode sum approach. 

The paper is organized as follows. In the next section, we briefly introduce the preliminaries of bulk reconstruction in AdS and discuss the existing dS results. The second section sets up the computation for boundary representations of scalars. The results are presented in the third section. In the fourth section, we turn to higher spin fields. We conclude with a summary of our results. We provide an appendix where we collate useful results on Weber-Schafheitlin type integrals. 

\section{Preliminaries}
\subsection{Bulk Reconstruction in AdS}
In this section, we briefly recall the details of bulk reconstruction  in a $d+1$-dimensional AdS. There are several equivalent routes to bulk reconstruction, we highlight the mode sum method which we have used in this paper. 

We work in the Poincare patch $(z,\bx), \bx = x_0,x_1,..x_d$. The metric is given by:
\begin{equation}
    ds^2= \frac{1}{z^2}(dz^2 + d\bx^2) 
\end{equation}

We want to find the boundary representation for a free scalar field $\phi$ of mass $m$.  The Klein-Gordon equation for a free scalar is given by:
\beq \label{kg}
(\Box  -m^2) \phi =0
\eeq
There are two linearly independent solutions:
\beq 
\phi(z,\bx) =\phi_n(z,\bx) +\phi_{nn}(z,\bx)
\eeq
where, as $z \to 0$ 
\beq
\phi_{n}(z,\bx) \sim z^\Delta,\, \phi_{nn}(z,\bx) \sim z^{d-\Delta}
\eeq

Here $\phi_n(z,\bx)$ are (linear combinations of) normalizable mode that vanishes at the boundary and $\phi_{nn}(z,x)$ is the non-normalizable mode that blows up at the boundary. We discard the non-normalizable mode and keep only the normalizable one.

We then have a mode expansion:
\beq \label{mode}
\phi(z,\bx ) = \int d^dk\, \frac{1}{N_{\bk}} f_{k}(z) e^{i \bk \cdot \bx}a_{\bk } + \text{hermitian conjugate}
\eeq

The AdS/CFT extrapolate dictionary \cite{Banks:1998dd,Harlow:2011ke} relates the bulk correlation functions of the $\phi$ with that of a scalar primary operator $\O$ in the boundary CFT:
\begin{align}
\notag \lim_{z\to 0}z^{n\Delta} &\langle\phi(z,\bx_1 )\phi(z,\bx_2 )...\phi(z,\bx_n )\rangle \\ \label{xp}&= \langle 0|\O(\bx_1 )\O(\bx_2 )..\O(\bx_n )|0\rangle 
\end{align} 
Substituting this in \eqref{xp2} we obtain boundary representations for the creation and annihilation operators:
\beq \label{anni}
a_{\bk}= \frac{N_{\bk}}{C_{\bk}}\int \O(\bx)\, d^dx 
\eeq
where 
\beq 
C_{\bk}= \lim_{z\to 0}z^{\Delta} f_{k }(z). 
\eeq

where $\Delta= \frac{d}{2}+ \sqrt{\frac{d^2}{4}+m^2}$ is the conformal dimension of $\O(\bx)$. Within correlators, we may then write:
\beq \label{xp2}
\lim_{z\to 0}z^{\Delta} \phi(z,\bx )=\O(\bx )
\eeq

Then substituting \eqref{anni} in \eqref{mode}, one obtains the boundary representation of the bulk scalar, which can be expressed in the form:

\beq 
\phi(z,\bx) = \int  d^dx^\prime\, K(z,\bx;{\bx^\prime})O({\bx^\prime})
\eeq
where 
\beq 
K(z,\bx ;{\bx^\prime})=\int d^dk\, f_{k }(z) e^{i \bk \cdot (\bx -\bx^\prime )}
\eeq 
is called the smearing function. This construction is  called the HKLL construction \cite{Hamilton:2005ju,Hamilton:2006az}. An alternative route to HKLL construction is via the construction of a spacelike Green function\cite{Hamilton:2005ju,Hamilton:2006az}. 

\subsection{Bulk Reconstruction in dS}

We consider a $d+1$-dimensional de Sitter metric and work in the flat slicing:
\beq \label{met}
ds^2= \frac{-d\eta^2 + d\x \cdot d\x}{\eta^2}; \qquad \x= (x_1,x_2,..x_d) 
\eeq 
Note that we have set Hubble constant $H=1$. This coordinate chart covers only the future space-like boundary which is at $\eta \to 0.$ 

Just like in AdS, the Klein-Gordon equation in dS has two independent solutions.  
\beq \label{norm}
\phi(\eta,\x) =c_\Delta \phi_{\Delta}(\eta,\x) + c_\-\phi_{d-\Delta}(\eta,\x)
\eeq
where $c_\+,c_\-$ are constants. 
The boundary behavior is given by:
\beq \label{fall}
\lim_{\eta \to 0}\phi_{\Delta}(\eta,\x) = \eta^\Delta \O_{\Delta}(\x),\, \phi_{d-\Delta}(\eta,\x) = \eta^{d-\Delta}\O_{d-\Delta}(\x)
\eeq
Here we take 
\begin{align}\notag
\Delta &= \frac{d}{2} + \sqrt{\frac{d^2}{4} - m^2}\quad  \text{if} \, \, \frac{d^2}{4} > m^2\\ \label{deltas}
&=\frac{d}{2} +i \sqrt{m^2-\frac{d^2}{4}  }\quad \text{if}\, \,\frac{d^2}{4} < m^2
\end{align}

$\O_\Delta(x),\O_{d-\Delta}(x)$ transform under de Sitter symmetries in the same way as a conformal primaries of conformal weight $\Delta, d-\Delta$ respectively. 

Unlike in the case of AdS, both the modes are normalizable and neither be discarded. We would therefore have to obtain boundary representations for both:

\begin{align} 
\label{smd}\phi_{\Delta}(\eta,x)= \int d^dx^\prime \,K_{\Delta}(\eta,x;x^\prime)\O_\Delta(x^\prime) \\
\label{smdd}\phi_{d-\Delta}(\eta,x)= \int d^dx^\prime\, K_{d-\Delta}(\eta,x;x^\prime)\O_{d-\Delta}(x^\prime)
\end{align}

The boundary representation of $\phi$ is then given by \eqref{norm}.

In \cite{Xiao:2014uea}, the smearing function was found for the case of a scalar field of mass $m^2> d^2/4$, which is a scalar field belonging to the principal series representation. This was done for the Bunch-Davies vacuum. The Wightman function and Green functions in de Sitter were used to read off the smearing function. The result obtained was the following: 
\begin{equation}\label{xiao}
\begin{aligned}   K_\+(\eta, \x, \x^\prime)=\frac{\Gamma\left(\Delta-\frac{d}{2}+1\right)  }{\Gamma(\Delta-d+1)}&\left(\frac{\eta^{2}-|\Delta \x|^{2}}{\eta}\right)^{\Delta-d}\\&\theta(\eta-|\Delta \x|)\\
K_\-\left(\eta,\x; \x^{\prime}\right)=\frac{\Gamma(d / 2-\+ +1)}{\Gamma(1 -\+)} &\left(\frac{\eta^2-|\+\x|^2}{\eta}\right)^{-\+}\\&\theta(\eta-|\Delta \x|)
\end{aligned}
\end{equation}

As was noted in that paper, the two smearing functions appearing here are exactly what one would obtain if one were to analytically continue the smearing functions for two scalars dual to conformal primaries of dimensions $\+,\-$. One might thus consider such analytic continuation to be an alternative prescription for constructing boundary representations.

However, the method used in the paper only applied to principal series scalars. In particular, divergences appear in the formulae for the smearing function in the case where $\+$ is either zero or a negative integer. This case is relevant for constructing boundary representations of higher spin fields as they can be recast as a multiplet of scalars. The mass of these scalars is zero for a spin-two field and negative integers for higher spins (there is no divergence in the case of spin-one but it too can not be handled by this method \cite{Sarkar:2014jia}.

As we will see in the next section, using the mode sum approach helps us go beyond these limitations and construct the smearing function for the general case of a bosonic field in de Sitter.

\section{Boundary Representation of light fields in de Sitter}

We will now set up the computation for boundary representations for a massive scalar field in de Sitter. We do not restrict ourselves to any range of mass here. We will be working in the flat slicing and use the metric \eqref{met}.

In the flat slicing, the Klein-Gordon equation 
\beq
\left(\square-m^{2}\right) \phi(\eta,\x)=0 \\
\eeq
translates to
\beq
\left(\partial_{\eta}^{2}+\frac{1-d}{\eta} \partial_{\eta}+\frac{m^{2}}{\eta^{2}}-\partial_{i} \partial^{i}\right) \phi(\eta, \x)=0
\eeq

As noted in the previous section, the solution will have the form given by the equations \eqref{norm} and \eqref{fall}. 

Substituting

\beq
\phi(\eta,\x)=f_k(\eta) e^{i \k \cdot \x}
\eeq
we get:

\beq
\left(\partial_{\eta}^{2}+\frac{1-d}{\eta} \partial_{\eta}+k^{2}+\frac{m^{2}}{\eta^{2}}\right) f_k(\eta)=0    
\eeq
where $k=|\vec{k}|$.
For non-integer $\Delta$ general solution to this equation is given by:

\begin{equation}
f_k(\eta) =b_{\+} \eta^{d / 2} J_{\Delta-d / 2}(k \eta)+b_{\-} \eta^{d / 2} J_{d/2-\Delta}(k \eta).
\end{equation}

where $J_{\mu}(\eta)$ is the Bessel function of first kind. $b_{\+},b_{\-}$ are constants to be fixed from boundary conditions.
For integer $\+$, the two Bessel functions are not linearly independent and the solution is given by: 

\begin{equation}
f_k(\eta) =c_{\+} \eta^{d / 2} J_{\Delta-d / 2}(k \eta)+c_{\-} \eta^{d / 2} Y_{\Delta-d / 2}(k \eta)
\end{equation}
$Y_{\Delta-d/2}(\eta)$ is the Bessel function of the second kind respectively. $c_{\+},c_{\-}$ are constants to be fixed from boundary conditions.

The asymptotic behaviors of the modes are given by: 
\begin{align*}
\lim_{\eta \to 0}\eta^{d/2}J_{\Delta-d/2}(k\eta) \sim \eta^\Delta\\
\lim_{\eta \to 0}\eta^{d/2}J_{d/2-\Delta}(k\eta) \sim \eta^{d-\Delta}\\
\lim_{\eta \to 0}\eta^{d/2}Y_{\Delta-d/2}(k\eta) \sim \eta^{d-\Delta}
\end{align*}

We can identify the modes corresponding to the two different fall-off behaviors. 

The boundary representation of $\phi$ will always feature two different smearing functions corresponding to these two modes, as in equations \eqref{smd} and \eqref{smdd}. When $\+$ is a non-integer, we construct the smearing function for \eqref{smdd} using $J_{d/2-\Delta}$, while for the case of integer $\+$, we will use $Y_\-$. As we will see, we get the same functional form in both cases, but the constant prefactors are different.

\subsection{Smearing function for $\phi_\+$}

First, we consider $\phi_\+$ and set up the construct of the corresponding smearing function. 

$\phi_\Delta$ can be mode expanded as: 
\beq \label{mode}
\phi_{\Delta}= \int d^d k \eta^{d / 2} J_{\Delta-d / 2}\left(k {\eta}\right)\left(e^{i \k \cdot \x} a^{\dagger}+e^{-i \k \cdot \x} a\right)
\eeq

Substituting the above in the extrapolate dictionary

\beq
\lim_{\eta \rightarrow 0} \phi_{\Delta}(x,\eta)=\eta^{\Delta} \O_{\Delta}(\x)
\eeq

and using the asymptotic expression
\beq \label{as}
\lim_{\eta \to 0}J_{\Delta-d/2}(k\eta) =  \frac{1}{\Gamma(\Delta-d / 2+1)}\left(\frac{k\eta}{2}\right)^{\Delta -d/2}
\eeq
we have:
\begin{equation}
 \frac{1}{\Gamma(\Delta-d / 2+1)}\left(\frac{k} {2}\right)^{\Delta-d / 2}( e^{i \k \cdot \x} a^{\d}+e^{-i \k \cdot \x} a)=\operatorname{\O}_\+(\x)
\end{equation}

Using inverse Fourier transform on the above equation we express the creation and annihilation operator in terms of a operator in the boundary CFT as:

\begin{equation}
\begin{aligned}
& a^{\d}=\int \Gamma\left(\Delta-\frac{d}{2}+1\right)\left(\frac{2}{k}\right)^{\Delta-d / 2} e^{-i k \cdot x^{\prime}} {\O}_\+\left(\x^{\prime}\right) d x^{\prime} \\
& a=\int \Gamma\left(\Delta-\frac{d}{2}+1\right)\left(\frac{2}{k}\right)^{\Delta-d / 2} e^{i k \cdot x^{\prime}} {\O}_\+\left(\x^{\prime}\right) d x^{\prime}
\end{aligned}
\end{equation} 

Substituting the above in \eqref{mode} gives us:
\beq
\phi_{\+}(\eta,\x)=\int K_\+\left( \eta,\x ; \x^{\prime}\right) \operatorname{\O}_\+\left(\x^{\prime}\right) d^{d} x^{\prime}
\eeq
with
\beq \label{smear}
\begin{gathered}
K_\+\left(\eta,\x; \x^{\prime}\right)=\Gamma(\Delta-d / 2+1) 2^{\Delta-d / 2} \int \eta^{d / 2} J_{\Delta-d / 2}(k \eta) \\ k^{d / 2-\Delta}
e^{i \k \cdot \Delta \x} d^{d} k 
\end{gathered}
\eeq

Here, $\Delta \vec x=\x-\x^{\prime}$ 

We can perform the angular integral in \eqref{smear} using \cite{Bhattacharjee:2022ehq} :
\beq \label{angle}
\int d^dk\, e^{i\k\cdot\+ \x} k^\mu f(k) =2^{d/2} \int dk\, \frac{k^{\mu-d/2}}{|\+ \x|^{d/2-1}}f(k)
\eeq

we obtain: 
\beq \label{int1}
\begin{aligned}
K_\+\left( \eta, \x; \x^\prime\right)=& 
 \frac{\Gamma(\Delta-d / 2+1) 2^{\Delta}}{|\Delta \x|^{d/2-1}}\\& \int_{0}^{\infty} \eta^ {d/2} k^{d-\Delta} J_{\Delta-d / 2}\left(k \eta\right) J_{\frac{d}{2}-1}(k |\Delta x|) d k
\end{aligned}
\eeq

This is a Weber-Schafheitlin type integral. Expressions for these integrals are available in all cases of interest to us, but the expressions are not always analytic. We have collected the relevant results on these integrals in the appendix. 

The smearing function has an analytic expression for principal series scalars of all masses when $d=1$. For lighter scalars, it has an analytic expression 
\begin{enumerate}
    \item for all values of mass when $d=1$
    \item for $m^2<d-1$ when $d\geq 2$.
\end{enumerate}

In this case, the expression for the smearing function is:

\begin{align}
\notag K_\+(\eta, \x, \x^\prime)&=\frac{\Gamma\left(\Delta-\frac{d}{2}+1\right)  }{\Gamma(\Delta-d+1)}\\& \label{sm0}\left(\frac{\eta^{2}-|\Delta \x|^{2}}{\eta}\right)^{\Delta-d}\theta(\eta-|\Delta \x|)
\end{align}

This follows from \eqref{analy1} together with the hypergeometric identity $ {}_2F_1{\left( a, b; a ; z \right )} = (1-z)^{-b}$. 
$\theta(\eta-|\Delta \x|) $ has appears in the expression because integral vanishes when $|\Delta \x|>\eta$.

Note that this matches exactly with the smearing function \eqref{xiao} obtained in \cite{Xiao:2014uea}. 

The smearing function is distributional for principal series scalars of all masses in $d>1$. For all lighter scalars, it is analytic in spatial dimensions $d\geq2$ for the mass range $d-1\leq m^2 \leq d^2/4$. We do not quote the full expression here, rather we refer the reader to expressions \eqref{dist1a} and \eqref{dist1a} in the appendix, with the following substitutions:
\begin{align*}
\lambda & \to \+-d\\
\mu &\to \Delta-d/2\\
\nu &\to d/2-1.
\end{align*}

\subsection{Smearing function for $\phi_\-$}

\subsubsection{Case of Non-Integer $\+$}
Following the same steps as before and using \eqref{as}, we obtain the following expression for the smearing function: 

\beq 
\begin{aligned}
K_\-\left(\eta,\x; \x^{\prime}\right)=\Gamma(d / 2-&\+ +1) 2^{d / 2-\Delta} \int d^{d} k \,\eta^{d / 2}   \\
&k^{\Delta-d/2}e^{i \k \cdot \Delta \x} J_{d / 2-\Delta}(k \eta)
\end{aligned}
\eeq

After performing the angular integral using \eqref{angle}, we arrive at:
\beq 
\begin{aligned}
K_\-\left(\eta,\x; \x^{\prime}\right)=&\frac{\Gamma(d / 2-\+ +1) 2^{d -\Delta}}{|\Delta \x|^{d/2-1}} \int d^{d} k \,\eta^{d / 2}   \\
&k^{\Delta} J_{d / 2-\Delta}(k \eta)J_{d / 2-1}(k |\Delta \x|)
\end{aligned}
\eeq

This is the same type of integral we encountered. For principal series scalars, it has an analytic expression for $d=1$. For lighter scalars, it has an analytic expression only in $d=1$ and that too for scalars of mass $m^2<d-1$. 

Using \eqref{analy1} and the hypergeometric identity ${}_2F_1(a,b;a;z)=(1-z)^{-b}$, this expression turns out to be:

\beq \label{sm1}
K_\-\left(\eta,\x; \x^{\prime}\right)=\frac{\Gamma(d / 2-\+ +1)}{\Gamma(1 -\+)} \left(\frac{\eta^2-|\+\x|^2}{\eta}\right)^{-\+}
\eeq

This matches with the expression \eqref{xiao} obtained in \cite{Xiao:2014uea}. 

The integral becomes distributional for principal series scalars when $d>1$. For lighter scalars, it is distributional for $d>1$ when the mass satisfies $d-1\leq m^2\leq d^2/4$.

In this case, the smearing function can be read off from \eqref{dist1} of the appendix, with the following substitutions:
\begin{align*}
\lambda & \to -\+\\
\mu &\to \Delta-d/2\\
\nu &\to d/2-1.
\end{align*}
\subsubsection{Case of Integer $\+$}
This only applies for scalars with $m^2<d^2/4$, since $\+$ is real in this case.
Again we follow the same steps as before.  Using the asymptotic expression
 \beq 
\lim_{\eta \to 0}Y_{\Delta-d/2}(k\eta) =-\frac{\pi}{\Gamma(\Delta-d/2)} \left(\frac{k\eta}{2}\right)^{d/2-\Delta }
\eeq

we arrive at the expression for the smearing function:
\begin{align}\notag
K_\-( \eta,\x; \x^\prime) =&-\frac{\pi}{\Gamma(\Delta-d/2)  2^{\Delta-d/2}} \\& \int d^d k  \eta^{d / 2} k^{\Delta-d / 2} Y_{\Delta-d / 2}(k \eta)\; e^{i \k\cdot \Delta \x} 
\end{align}

 After performing the angular integral using \eqref{angle}, this reduces to:
\begin{align}\notag
K_\-( \eta,\x; \x^\prime)=&\frac{\pi}{\Gamma(\Delta-d/2) 2^{\Delta-d}|\Delta \x|^{d / 2-1}}
\eta^{d / 2} \\ \label{int2} & \int_{0}^{\infty} dk\, k^{\Delta} Y_{\Delta-d/ 2}\left(k \eta\right) J_{d / 2-1}(k |\Delta \x|)  
\end{align}

This is another Weber-Schafheitlin type integral. This has an analytic expression in all cases of our interest. Using \eqref{id} and \eqref{analy2} along with the hypergeometric identity $ {}_2F_1{\left( a, b; b ; z \right )} = (1-z)^{-a}$, we obtain the following expression 

 \begin{align}\notag \label{sm2}
K_\-(\eta,\x; \x^\prime) =& - \frac{(4 \pi)^{\frac{d}{2}}\Gamma({\Delta})}{\Gamma(\Delta-d/2)}\sin\left(\frac{\pi}{2}(d-1)\right) \\& \left(\frac{\eta^2-|\Delta \x|^2}{\eta}\right)^{-\Delta}\theta(\eta-|\Delta \x|)
\end{align}

 $\sin\left(\frac{\pi}{2}(d-1)\right)=0$ vanishes when $d$ is odd and equals unity when $d$ is even. 

 We thus have the intriguing result that the smearing function vanishes for integer $\Delta$ in even spacetime dimensions. This requires further investigation. 

 For odd dimensional space-times, we note that the functional form of the smearing function matches with the non-integer case \eqref{sm1}, but the coefficient is different. In particular, the gamma functions are no longer divergent when $\+$ is a positive integer, which is the case of interest for higher spin gauge fields!

\subsection{ Smearing function for different vacua}
So far, we have not fixed the constants $c_\Delta, c_\-$ that appear in \eqref{norm}. These are fixed through boundary conditions, or equivalently, choice of vacuum. Here we fix the boundary conditions for light fields in the case where $\+$ is an integer. The extension to the non-integer case is straightforward. 

In inflationary applications, one typically chooses the Bunch-Davies vacuum. This corresponds to the solution:
\beq
f^{BD}_k(\eta) = \eta^{d / 2}H^{(2)}_{\Delta-d / 2}(k \eta) 
\eeq

where $H_{\nu}^{(2)}$ is the Hankel function of second kind. Using the well-known relation between Hankel and Bessel functions:
\beq
H_{\nu}^{(2)}(x) = J_{\nu}(x) -i Y_{\nu}(x)
\eeq
\\
we have $c_\+= 1 , c_\-= -i$ for Bunch-Davies vacuum.

Then we have:
 \begin{align} \notag
\phi(\eta,\x)=&  \int d^d x^\prime K_\+(\eta,\x; \x^\prime)\O_\+(\x^\prime)  \\
& -i \int d^d x^\prime K_\-(\eta,\x; \x^\prime)\O_\-(\x^\prime)
\end{align}

One can consider a more general family of vacua which are invariant under the de Sitter group. These are the $\alpha$-vacua\cite{Mottola:1984ar,Allen:1985ux},  parametrized by a single positive real parameter $\alpha$. They are given by:
\beq \label{alpha}
f^{\alpha}_k(\eta) =\cosh \alpha f^{BD}_k(\eta) + \sinh \alpha \left(f^{BD}_k(\eta)\right)^*.
\eeq

For an $\alpha$-vacuum, the constants then are:
\begin{align*}
  c_\+&= \cosh \alpha + \sinh \alpha \\
  c_\- &= i \left(\sinh \alpha -\cosh \alpha\right)
\end{align*}
Here we used the fact that for real $\mu$, $J_\mu(\Bar{z})=\overline{J_\mu}(z),Y_\mu(\Bar{z})=\overline{Y_\mu}(z)$.
Hence the boundary representation of a bulk field in an $\alpha$-vacuum is given by: 
\begin{align} \notag
\phi&(\eta,\x)=\left(\cosh \alpha +\sinh \alpha\right) \int d^d x^\prime K_\+(\eta,\x; \x^\prime)\O_\+(\x^\prime)  \\
& + i \left(\sinh \alpha -\cosh \alpha\right)  \int d^d x^\prime K_\-(\eta,\x; \x^\prime)\O_\-(\x^\prime)
\end{align}

 One can do the same for the case of the non-integer $\+$ by using the identity: 
 \beq 
Y_\mu(x)= \frac{\cos (\mu \pi) J_\mu(x) - J_{-\mu}(x)}{\sin (\mu \pi)}
 \eeq


\section{Boundary representation for higher spin fields in de Sitter}
We now turn to the computation of boundary representations for gauge fields with general integer spin.

First, let us consider a spin-1 field. It can be recast as a multiplet of scalar fields of equal mass. We describe this construction by following the analogous construction in AdS\cite{Kabat:2012hp}.

Let us consider a spin-1 field satisfying the source-free Maxwell equation:
\beq 
\nabla_\mu F^{\mu }=0.
\eeq

In the holographic gauge
\beq \label{gauge}
A_\eta(\eta,\x) =0.
\eeq

the equations of motion simplify: 
\beq 
\partial_\eta (\partial_i A^j)=0
\eeq

The residual gauge symmetry left by \eqref{gauge}:
\beq 
A_i(\eta,\x) = A_i(\eta,\x) +\partial_i\lambda(\x)
\eeq
allows one to fix
\beq
\partial_i A^i =0
\eeq

The Maxwell's equations then simplify: 
\beq 
\eta^{d-3}\partial_\eta \left(\eta^{3-d}\partial_\eta A_i(\eta,\x)\right) + \partial_i\partial^i A_i(\eta,\x)=0.
\eeq

Substituting $\phi_i(\eta,\x) =\eta A_i (\eta,x)$ gives us:
\beq
\left(\partial_{\eta}^{2}+\frac{1-d}{\eta} \partial_{\eta}+\frac{d-1}{\eta^{2}}-\partial_{i} \partial^{i}\right) \phi_i(\eta, \x)=0
\eeq

Thus the source-free Maxwell's equations for a spin-1 field reduce to a set of $d$ Klein-Gordon equations for a free scalar of mass $m^2=d-1$. 

The approach of Wightman functions followed in \cite{Xiao:2014uea} was found to be inadequate for spin-1 case \cite{Sarkar:2014dma}. Our results provide a clue as to why this was the case. In $d>1$, the smearing function $K_\Delta(\eta, |\+\x|)$ becomes distributional for $m^2<d-1$ (this is a strict inequality). Hence one needs to use \eqref{dist1} if $\+=d/2 +\sqrt{d^2/4-d+1}$ is a non-integer or \eqref{dist1a} if $\+$ is an integer (with the appropriate substitutions). 

$K_\-(\eta, |\+\x|)$ will also be non analytic for the case of non-integer  $\+$ except for $d=1$. Hence \eqref{dist1} will apply(with the appropriate substitutions). When $\+$ is an integer, \eqref{sm2} gives the correct smearing function. Curiously, $\+$ is an integer only for odd-dimensional spacetimes, which means the issue of the vanishing of smearing functions never arises for spin-1 fields.

We now consider general integer spin fields. Once again, one can recast the fields as a multiplet of scalars using a similar construction. We outline it briefly following \cite{Sarkar:2014dma}, readers can consult that paper for details.

Consider a massless gauge field with integer spin $s$, It is represented by a totally symmetric tensor $F_{A_1...A_s}$ of rank $s$ which satisfies double-tracelessness: 
\beq 
{F^{AB}}_{ABA_5...A_s}=0
\eeq
This linear equation of the field is\cite{Vasiliev:1995dn}:
\beq\label{eom2}
\begin{aligned}
&\nabla_B\nabla^B F_{A_1..A_s}-s\nabla_B\nabla_{A_1}{F^{B}}_{A_2..A_s}+\frac{1}{2}s(s-1)\\&\nabla_{A_1}\nabla_{A_2}{F^{B}}_{BA_3..A_s}-2(s-1)(s+d-2)F_{A_1..A_s}=0.
\end{aligned}
\eeq

The holographic gauge is the one in which the $\eta$-components are taken to vanish:
\beq \label{gauge}
F_{\eta...\eta} =F_{ \mu_1 \eta...\eta}=\dots=F_{ \mu_1.. \mu_{s-1}\eta}=0.
\eeq
Here $\mu$ is the index used for $A$ other than $\eta$.  
One can further choose to impose
\beq 
{F^ \nu}_{ \nu \mu_3... \mu_s}=0, \partial_ \nu{F^ \nu}_{ \nu \mu_3... \mu_s}=0
\eeq
which fixes residual gauge conditions.

 Now we define a multiplet of scalars by:
 \beq 
\phi_{ \mu_1.. \mu_s}=\eta^s F_{ \mu_1.. \mu_s}
 \eeq

We find that the equations of motion \eqref{eom2} simplify to 
 
\beq
\left(\partial_{\eta}^{2}+\frac{1-d}{\eta} \partial_{\eta}+\frac{m^2}{\eta^{2}}-\partial_{\mu} \partial^{\mu}\right) \phi_{ \mu_1.. \mu_s}(\eta, \x)=0
\eeq
 where $m^2=(2-s)(s+d-2)$. 

Thus the equations of motion simplifies to a set of scalar fields of mass $m^2=(2-s)(s+d-2)$, or equivalently $\+=s+d-2$. 

As we saw earlier, smearing functions $K_\-(\eta, |\+\x|)$ corresponding to these integer valued $\Delta$ were found to be divergent in \cite{Xiao:2014uea}. We now see that the appropriate formula for this case is \eqref{sm2}, which does not diverge. $K_\+(\eta, |\+\x|)$ will be analytic in this case, and be given by \eqref{sm0}. Thus we always get nice analytic expressions for gauge fields with spin greater than one. However, the smearing function vanishes in even spacetime dimensions.

\section{Discussions}

The bulk reconstruction program for de Sitter spacetime was started in \cite{Xiao:2014uea} and boundary representations were obtained for principal series scalars. But for scalars with ($m^2 \leq d-1$) as well as higher spin fields, the construction ran into issues. Especially for higher spin fields, divergences were encountered. 

In this paper, we obtained boundary representations for all scalars and massless higher spin fields. We used the mode sum approach. We recovered the results of \cite{Xiao:2014uea} for certain ranges of mass and dimension. But were able to go extend the construction and resolve the issues of the divergences. For integer $\+$, it turned out that the smearing function was different from the one previously derived and free from divergences. This helped solve the outstanding issues for fields of spin 2 or beyond. 

We found that the smearing functions in de Sitter are given by Weber-Schafheitlin type integrals. Depending on the values of the parameters involved, these integrals may give an analytic or distributional expression. In our case, this translated to the result that for some values of dimension, mass and spin, the smearing function has a distributional nature. The fact that smearing functions can become distributional in pure dS is quite interesting--in asymptotically AdS spacetimes the smearing function only becomes distributional when event horizons are present. Our result also indicates that analytically continuing the smearing functions from AdS may not always give the correct answer(a similar point was made in the recent preprint \cite{Doi:2024nty}).  

We also considered the possibility of different boundary conditions, which were not considered in previous literature. Only Bunch-Davies initial conditions had been considered previously. We extended the construction of boundary representations from Bunch-Davies vacuum to general $\alpha$-vacua.  

A curious result that we obtained was the vanishing of smearing functions for integer $\+$ in even spacetime dimensions. This happens in the physically interesting case of higher spin fields. It is not entirely clear if this is an issue with the mode-sum approach and can be resolved using a different approach. One is reminded of the fact that in AdS, the Green function approach to bulk reconstruction fails in a similar way by giving vanishing smearing function in odd dimensions \cite{Hamilton:2006az,Heemskerk:2012mn, Bhattacharjee:2022ehq}. It will be interesting to explore this result further.

\begin{acknowledgments}
We would like to thank Michal Wrochna for the helpful discussions. NK is supported by SERB Start-up Research Grant SRG/2022/000970.
\end{acknowledgments}

\appendix* \section{Weber-Schafheitlin Type Integrals}
Weber--Schafheitlin type integrals \cite{watson1922treatise} have been known for a long time and have appeared in physical applications before\cite{richard2009new}. 
In this paper, we encounter this type of integral. 

The first one, which we encounter in \eqref{int1} and \eqref{int2} is of the type 
\beq
\int_{0}^{\infty} J_{\nu}(\alpha t) J_{\mu}(\beta t) t^{-\lambda} \,dt
\eeq

This integral has an analytic expression when $\operatorname{Re} \lambda>-1$. In this case, the result is \cite{Gradshteyn:1943cpj}
\beq \label{analy1}
\begin{aligned}
\int_{0}^{\infty} &J_{\nu}(\alpha t) J_{\mu}(\beta t) t^{-\lambda}\, dt = \frac{\alpha^{\nu} \Gamma{\frac{\nu+\mu-\lambda+1}{2}}}{2^{\lambda}\beta^{\nu-\lambda+1} \Gamma{\frac{-\nu+\mu+\lambda+1}{2} \Gamma{(\nu +1)}}}
\\ & \qquad F{\left( \frac{\nu+\mu-\lambda+1}{2} , \frac{\nu-\mu-\lambda+1}{2} ; \nu+1 ; \frac{\alpha^2}{\beta^2} \right )}\\ &\qquad \qquad 
\left [ \operatorname{Re}(\nu+\mu-\lambda+1) > 0, \; \;  0<\alpha < \beta \right]\\
& \qquad \qquad \qquad= \text{Above expression with } \nu\leftrightarrow \mu\\  & \qquad \qquad
\left [ \operatorname{Re}(\nu+\mu-\lambda+1) > 0, \; \;  0 < \beta <\alpha \right]
\end{aligned}
\eeq 
The expression vanishes when either $\nu+\mu-\lambda+1$ or $\nu-\mu-\lambda+1$ is an integer.

For the case $\operatorname{Re} \lambda\geq-1$, the expression becomes distributional. There have been several extensions beyond this limit \cite{salamon1979limits,miroshin2001asymptotic,kellendonk2009weber}. The most general result was derived in \cite{wrochna2010weber}, and is quoted below. Note that the result is to be interpreted as a distribution:

Case 1: For $\lambda\notin\Z$,
\begin{align}\notag 
&\int_{0}^{\infty} J_{\nu}(\alpha t) J_{\mu}(\beta t) t^{-\lambda} \, dt =  \\ \notag 
&-\frac{2^{-\lambda}}{\sin(\pi\lambda)} \bigg[ \bigg\{ \left(\frac{\beta}{\alpha}-1\right)_{-}^{\lambda}\sin\left(\pi \frac{1+\lambda-\mu+\nu}{2} \right)  \\   \notag  
&+ \left(\frac{\beta}{\alpha}-1\right)_{+}^{\lambda}\sin\left(\pi \frac{1-\lambda-\mu+\nu}{2} \right) \bigg\} \frac{\left(\frac{\beta}{\alpha}\right)^{-1-\lambda-\nu}}{\left(1+\frac{\beta}{\alpha}\right)^{-\lambda}}  \\\notag  
& {}_{2}F_1^{\rm I}\bigg(\frac{1+\lambda+\mu+\nu}{2}, \frac{1+\lambda-\mu+\nu}{2};\lambda+1; \\ \notag &1-\left(\frac{\beta}{\alpha}\right)^{-2}\bigg)
- \sin\left(\pi \frac{1-\lambda-\mu+\nu}{2} \right)\left(\frac{\beta}{\alpha}\right)^{-1+\lambda-\nu} \\ \notag & \frac{\Gamma\left(  \frac{1-\lambda+\mu+\nu}{2}\right)\Gamma\left(  \frac{1-\lambda-\mu+\nu}{2}\right)}{\Gamma\left(  \frac{1+\lambda+\mu+\nu}{2}\right)\Gamma\left(  \frac{1+\lambda-\mu+\nu}{2}\right)} \ {}_{2}F_1^{\rm I}\bigg(  \frac{1-\lambda+\mu+\nu}{2},\\ \label{dist1a} & \mspace{30mu}\frac{1-\lambda-\mu+\nu}{2};1-\lambda;1-\left(\frac{\beta}{\alpha}\right)^{-2}\bigg) \bigg].
\end{align}
 Case 2: For  $\lambda\in\Z$ and $\frac{1-\lambda\pm\mu+\nu}{2}\notin\Z$,
\begin{align}\notag
&\int_{0}^{\infty} J_{\nu}(\alpha t) J_{\mu}(\beta t) t^{-\lambda} \, dt =
 \bigg[ \bigg\{ \sin\left( \pi \frac{1-\lambda-\mu+\nu}{2} \right)\\ \notag 
&\left[\left(\frac{\beta}{\alpha}-1\right)^{\lambda}_- + (-1)^{-\lambda} \left(\frac{\beta}{\alpha}-1\right)^{\lambda}_+ \right] \\  \notag 
&+\cos\left( \pi \frac{1-\lambda+\mu-\nu}{2} \right) (-1)^{-\lambda} \pi \frac{\delta^{(-\lambda-1)}\left(1-\frac{\beta}{\alpha}\right)}{(-\lambda-1)!} \bigg\} \\  \notag 
&\times \left(\frac{\beta}{\alpha}\right)^{-1-\lambda-\nu} \left(1+\frac{\beta}{\alpha}\right)^{\lambda} S_{\mu,\nu,\lambda}\left(1-\left(\frac{\beta}{\alpha}\right)^{-2}\right) + \\ \notag 
&(-1)^{\lambda} \frac{\Gamma\left(- \frac{1-\lambda+\mu+\nu}{2}\right) \Gamma\left( \frac{1-\lambda-\mu+\nu}{2}\right)}{\Gamma\left( \frac{1+\lambda+\mu+\nu}{2}\right) \Gamma\left( \frac{1+\lambda-\mu+\nu}{2}\right)} \left(\frac{\beta}{\alpha}\right)^{-1+\lambda-\nu} \\  \notag 
&\bigg\{ \sin\left( \pi \frac{1-\lambda-\mu+\nu}{2} \right) T_{\mu,\nu,\lambda}\left(1-\left(\frac{\beta}{\alpha}\right)^{-2}\right) \\   \notag 
&- \bigg[\sin\left( \pi \frac{1-\lambda-\mu+\nu}{2} \right) \log\bigg(\left(\frac{\beta}{\alpha}\right)^{-2} \left(\frac{\beta}{\alpha}+1\right) \\  \notag 
& \left|\frac{\beta}{\alpha}-1\right|\bigg) +\cos\left( \pi \frac{1-\lambda-\mu+\nu}{2} \right) \pi \theta\left(\frac{\beta}{\alpha}-1\right)\bigg] \\ \notag 
& {}_{2}F_1^{\rm I}\bigg( \frac{1-\lambda+\mu+\nu}{2}, \frac{1-\lambda-\mu+\nu}{2};1- \lambda; \\ &\label{dist1} 1-\left(\frac{\beta}{\alpha}\right)^{-2} \bigg) \bigg\} \bigg]\frac{1}{2^{\lambda}\pi}.
\end{align}

where ${}_{2}F_1^{\rm I}(a,b;c;z)=\frac{{}_{2}F_1(a,b;c;z)}{\Gamma(c+1)}$
 
 $S_{\mu,\nu,\lambda}, T_{\mu,\nu,\lambda}$ are defined for $\left|z\right|<1$ by
\[
S_{\mu,\nu,\lambda}(z):=\sum_{k=0}^{-\lambda-1}\frac{\left({  \frac{1+\lambda+\mu+\nu}{2}}\right)_k \left({  \frac{1+\lambda-\mu+\nu}{2}}\right)_k}{(1+\lambda)_k k!}z^k,
\]
\begin{align*}
&T_{\mu,\nu,\lambda}(z)=\sum_{k=0}^{\infty}\frac{\left({ \frac{1-\lambda+\mu+\nu}{2}}\right)_k\left({  \frac{1-\lambda-\mu+\nu}{2}}\right)_k}{(k-\lambda)!k!}z^k \\& \times \ \left\{\psi(k+1)+\psi(-\lambda+k+1) \right. \\& \left.-\psi\left({  \frac{1-\lambda+\mu+\nu}{2}}+k\right) -\psi\left({  \frac{1-\lambda-\mu+\nu}{2}}+k\right)\right\},
\end{align*}
with $\psi(y):=\frac{d}{d y}\Gamma(y)/\Gamma(y)$.
Further, for $\operatorname{Re} \gamma>-1$:
\[
(x-1)^{\gamma}_{-}:=\begin{cases}|x-1|^{\gamma},& x<1\\ 0, & x\geq1 \end{cases},
\]
\[
(x-1)^{\gamma}_{+}:=\begin{cases}0,& x<1\\ (x-1)^{\gamma}, & x\geq1 \end{cases},
\]

The condition for \eqref{dist1}and \eqref{dist1a} to hold is $\operatorname{Re}(\nu-\lambda+1)>|\mu|$.
The second Weber-Schafheitlin type integral we encounter is in \eqref{int2}. This is of the type: 
\beq 
\int_{0}^{\infty} x^{-\lambda} Y_{\mu}(a x) J_{\nu}(bx)\, dx
\eeq
 To evaluate this, we first use the identity  \cite{Gradshteyn:1943cpj}:
\begin{align}\notag
\int_{0}^{\infty} x^{-\lambda} Y_{\mu}(a x) J_{\nu}(bx)\, dx =\frac{2}{\pi} \sin{\frac{\pi(\nu-\mu-\lambda)}{2}} \\ \label{id}
\int_{0}^{\infty} x^{-\lambda} K_{\mu}(a x) I_{\nu}(bx)\, dx
\end{align}

The integral on the right-hand side is given by \cite{}:

\begin{align}\notag
\int_{0}^{\infty} & x^{-\lambda} K_{\mu}(a x) I_{\nu}(bx)\, dx =  \frac{b^\nu \Gamma {\left( \frac{1-\lambda+\mu+\nu}{2}\right)} \Gamma{\left( \frac{1-\lambda-\mu+\nu}{2}\right)}}{2^{(\lambda+1)} \Gamma(\nu+1) a^{-\lambda+\nu+1}} \\ \notag &{}
F{\left (\frac{1-\lambda+\mu+\nu}{2} ,\; \frac{1-\lambda-\mu+\nu}{2}; \; \nu+1; \frac{b^2}{a^2}  \right )} \\ \label{analy2} & \qquad  
\left[ a>b,  \operatorname{Re}(\nu -\lambda+1 \pm \mu) >0\right]
\end{align}

 \bibliography{dsbulk}
 
\end{document}